\documentclass[11pt]{article}

\usepackage{amssymb}

\begin{document}

\title{Moyal's Characteristic Function, the Density Matrix and von Neumann's Idempotent }
\author{B. J. Hiley.}
\date{TPRU, Birkbeck, University of London, Malet Street,\\ London WC1E 7HX.\\ \vspace{0.2cm}}
\maketitle

\begin{abstract}
In the Wigner-Moyal approach to quantum mechanics, we show that Moyal's starting point, the characteristic function \\$M(\tau,\theta)=\int \psi^{*}(x)e^{i(\tau {\hat p}+\theta{\hat x})}\psi(x)dx$,
  is essentially the primitive idempotent used by von Neumann in his classic paper ``Die Eindeutigkeit der Schr\"{o}dingerschen Operatoren''.  This paper provides the original proof of the Stone-von Neumann equation.  Thus the mathematical structure Moyal develops is simply a re-expression of what is at the heart of quantum mechanics and reproduces exactly the results of the quantum formalism.  The ``distribution function'' $F(X,P,t)$ is simply the quantum mechanical density matrix expressed in an $( X,P)$-representation, where $X$ and $P$ are the mean co-ordinates of a cell structure in phase space.  The whole approach therefore clearly has little to do with classical statistical theories but is a consequence of a non-commutative nature of the theory.
 \end{abstract}

 %Section 1

\section{Introduction.}
The Wigner-Moyal approach has its origins in  early attempt of Wigner \cite{wig} to  find quantum corrections to the statistical properties of thermodynamic quantum systems.   The major contribution to the development of the formalism that will be the subject of this paper came from the seminal paper of Moyal \cite{moy}, the aim of which was to ask whether the Hilbert space formalism of quantum mechanics disguised what is essentially a statistical theory based on non-commutative functions. To explore this possibility, Moyal proposed that we start from one of the standard tools of statistical theory, namely, the characteristic function.  This was not formed in the usual way but instead begins by introducing the operator function 
\begin{eqnarray}
{\hat M}(\tau,\theta)=e^{i(\tau{\hat p}+\theta{\hat x})} 
\end{eqnarray}									%eqn (1)
where ${\hat x}$ and ${\hat p}$ are operators. He then forms the expectation value
\begin{eqnarray}
M(\tau,\theta)=\int \psi^{*}(x,t) e^{i(\tau {\hat p}+\theta{\hat x})}\psi(x,t)dx
\end{eqnarray}									%eqn (2)
which he then treats as the characteristic function. Moyal shows that this function can be written in the form
\begin{eqnarray}
M(\tau,\theta)=\int\psi^{*}(x-\tau/2)e^{i\theta x}\psi(x+\tau/2)dx
\end{eqnarray}									%eqn (3)	
 In a conventional statistical theory, the probability distribution function is the Fourier transform of the characteristic function. Thus
 \begin{eqnarray}
F(x,p,t)=\frac{1}{2\pi^{2}}\int M(\tau,\theta)e^{i(\tau p+\theta x)}d\tau d\theta										%eqn (4)
\end{eqnarray}
Here $x$ and $p$ are continuous variables in some, as yet, unspecified phase space.  Moyal then treats $F(x,p,t)$ as a probability distribution function and using the standard methods of statistics to find the  time development of this distribution. 

For a particle in one dimension with Hamiltonian $ H= p^{2}/2m +V$, this time development can be written as
\begin{eqnarray}
\frac{\partial F(x,p,t)}{\partial t} + p/m\frac{\partial F(x,p,t)}{\partial x}=
\int J(x,p-p^{\prime})F(x,p^{\prime},t)dp^{\prime}
\end{eqnarray}									%eqn (5)
where 
\begin{eqnarray}
J(x,p-p^{\prime})=\int[V(x-y/2)-V(x+y/2)]e^{-i(p-p^{\prime})y}dy
\nonumber
\end{eqnarray}
The fact that we start from a characteristic function and that we are working in an $(x,p)$ phase space tends to imply that we are working with a classical statistical theory.
However exactly the same equation (5) can be derived starting with the density operator satisfying the standard quantum mechanical Liouville operator equation,
\begin{eqnarray}
i\hbar\frac{\partial{\hat\rho}}{\partial t} +[{\hat\rho},{\hat H}]=0
\end{eqnarray}									%eqn (5)
This was shown originally by Takabayasi \cite{tak} and extended in a novel way in Bohm and Hiley \cite{boh81}. This approach enabled  us to extended these ideas to the Dirac equation \cite{boh83}. 

How is this possible?  The two approaches seem to start from two entirely different standpoints.  One uses classical statistical methods, while the other uses the quantum formalism, yet they arrive at the same time evolution equation.   The purpose of this paper is to show that these are not two different starting points.  They are, in fact, the same starting point.  

What I will show is that the so called `characteristic function'  of equation (2)
can be derived from the primitive idempotent that von Neumann \cite{von} uses  in his original proof of what is now called the Stone-von Neumann theorem.  This theorem shows that the Schr\"{o}dinger representation is irreducible and unique up to a unitary equivalence.  What this result shows is that  the mathematics developed by Moyal emerges from the very heart of the quantum formalism.  In consequence  the function $F(x,p,t)$ is {\em not} a probability distribution function but a specific representation of the quantum mechanical density operator.  In fact it is a density matrix in a specific $(x,p)$-representation and this explains why it can take negative values.  There is no reason why the density matrix has to be positive.

The relation between the Wigner distribution and the density matrix has already been discussed in a little known paper by Baker \cite{bak}. However the method we use here is different from his.  He concentrates on the quasi-probability distribution where as I will adopt a more algebraic approach, focussing on idempotents.  This not only gives a different insight into the relation between the two approaches, it further justifies the relation between the between the Wigner-Moyal approach and the Bohm approach that was initially exploited in Brown and Hiley \cite{bro} and later developed in more detail by Hiley \cite{hil03}.

In section 2 we show the essential steps used by Moyal to arrive at equation 5.  In section 3 we show how this equation can be derived starting from the density operator.  In section 4 we introduce the idempotent that forms the basis of von Neumann's work and then show how it is equivalent to the characteristic function used by Moyal.

% Section 2

\section{Moyal's Quantum Dynamics.}

Let us first begin by sketching the essential steps used by Moyal in his derivation of equation (5).  Moyal starts by assuming that $F(x,p,t)$ is a quasi-probability distribution function whose time development is described by the standard Green's function equation
\begin{eqnarray}
F(p,x,t)=\int K(p,x|p_{0},x_{0},t-t_{0})F_{0}(p_{0},x_{0},t_{0})dp_{0}dx_{0}\nonumber
\end{eqnarray}
This equation can be expressed in the intro-differential form
\begin{eqnarray}
\frac{\partial F(p,x,t)}{\partial t}=\int S(p,x|\eta,\xi)F(\eta,\xi,t)d\eta d\xi\nonumber
\end{eqnarray}
where $S(p,x|\eta.\xi)$ is the derivative of the kernel $K(p,x|p_{0},x_{0},t-t_{0})$.
However the key to Moyal's approach is to work with the characteristic function defined in equations (1) and (7).  Thus if $M(\tau,\theta,t_{0})$ is the characteristic function at $t_{0}$ then
\begin{eqnarray}
M(\tau,\theta,t_{0})=\int e^{i(\tau\eta+\theta\xi)}F_{0}(\eta,\xi,t_{0})d\eta d\xi\nonumber
\end{eqnarray}
while the characteristic function at $t$ is
\begin{eqnarray}
M(\tau,\theta,t)=\int e^{i(\tau\eta+\theta\xi)}\Lambda(\tau,\theta|\eta,\xi,t-t_{0})F_{0}(\eta,\xi,t_{0})d\eta d\xi\nonumber
\end{eqnarray}
with
\begin{eqnarray}
\Lambda(\tau,\theta|\eta,\xi,t-t_{0})=\int e^{i[\tau(p-\eta)+\theta(x-\xi)]}K(p,x|\eta,\xi,t-t_{0})dpdx\nonumber
\end{eqnarray}
Furthermore
\begin{eqnarray}
\frac{\partial M}{\partial t}=\int L(\tau,\theta|\eta,\xi)e^{i(\tau\eta+\theta\xi)}F(\eta,\xi, t)d\eta d\xi \nonumber
\end{eqnarray}
and
\begin{eqnarray}
S(p,x|\eta,\xi)=\int L(\tau,\theta|\eta,\xi)e^{i[\tau(\eta-p)+\theta(\xi-x)]}d\tau d\theta\nonumber
\end{eqnarray}
Thus we need to determine $L(\tau,\theta|\eta,\xi)$ from $\partial M/\partial t$ to determine $S(p,x|\eta,\xi)$ which in turn can be used to determine $\partial F/\partial t$.  This will give us the time development equation (5).

The vital step then is to derive the time derivative of the characteristic function, namely,  $\partial M/\partial t$.  What Moyal does is to assume that this function satisfies a quantum Liouville equation
\begin{eqnarray}
i\hbar\frac{\partial {\hat M}}{\partial t}=[{\hat H},{\hat M}]\nonumber
\end{eqnarray}
where ${\hat M}=e^{i(\tau {\hat p}+\theta {\hat x})}$ and ${\hat H}$ is the Hamiltonian written in operator form.  Of course it is immediately evident that ${\hat M}$ is an element of the Heisenberg group.  This suggests that basically Moyal is starting from quantum mechanics but we should be cautious with this conclusion because the Heisenberg group can be derived from a purely classical wave theory as was clearly demonstrated in Moran and Manton \cite{mor}.  However the use of  a quantum equation of motion suggests that the Moyal approach may be pure quantum mechanics albeit in a different mathematical form.  The question of why this characteristic function should satisfy this equation is not discussed by Moyal.

% Section 3

\section{Moyal Dynamics Deduced from the Density Matrix.}

Let us now show how equation (5) follows directly from the standard quantum mechanical equation of the time development density operator (6).  In the $x$-representation this can be written in the form
\begin{eqnarray}
i\frac{\partial  \langle x|{\hat \rho}|x^{\prime}\rangle}{\partial t}=
\langle x |[{\hat H},{\hat \rho}]|x^{\prime}\rangle \nonumber
\end{eqnarray}					
which gives
\begin{eqnarray}
i\frac{\partial\rho(x,x^{\prime})}{\partial t}+\frac{1}{2m}\left(\frac{\partial^{2}}{\partial x^{2}}-\frac{\partial^{2}}
{\partial x^{\prime 2}}\right)\rho(x,x^{\prime})=[V(x)-V(x^{\prime})]\rho(x,x^{\prime})
\end{eqnarray}									% eqn (7)
The density matrix $\rho(x,x^{\prime})$ can be regarded as a two point function in configuration space.  By writing this density matrix as
\begin{eqnarray}
\rho(x,x^{\prime},t)=\psi(x,t)\psi^{*}(x^{\prime},t)=\frac{1}{2\pi}\int\phi(p,t)e^{ixp}\phi^{*}(p^{\prime},t)e^{-ix^{\prime}p^{\prime}}dpdp^{\prime}\nonumber
\end{eqnarray}
and introducing new co-ordinates
\begin{eqnarray}
X=(x^{\prime}+x)/2;  \hspace{0.3cm}  \tau=x^{\prime}-x;  \hspace{0.3cm}
\mbox{and}  \hspace{0.3cm}   P=(p^{\prime}+p)/2;  \hspace{0.3cm}  \theta=p^{\prime}-p  
\end{eqnarray}										%eqn (8)
the density matrix can be transformed into
\begin{eqnarray}
\rho(X,\tau,t)=\frac{1}{2\pi}\int\phi(P-\theta/2,t)e^{i\theta X}\phi(P+\theta/2,t)e^{i\tau P}dPd\theta  \nonumber
\end{eqnarray}
We can now write this equation in the compact form
\begin{eqnarray}
\rho(X,\tau, t)=\int F(X,P,t)e^{i\tau P}dP
\end{eqnarray}										%eqn (9)
where
\begin{eqnarray}
F(X,P,t)=\frac{1}{2\pi}\int\phi^{*}(P-\theta/2,t)e^{iX\theta,t}\phi(P+\theta/2,t)d\theta
\nonumber
\end{eqnarray}
Or taking the inverse Fourier transform we have
\begin{eqnarray}
F(X,P,t)=\frac{1}{2\pi}\int\psi^{*}(X-\tau/2,t)e^{iP\tau}\psi(X+\tau/2,t)d\tau
\nonumber
\end{eqnarray}
which, of course, is just the Wigner distribution \cite{wig}.   If we now change variables in equation (7) and use equation (9), we are led immediately to equation (5).

 From this derivation of the function $F(x,p,t)$ above, it should be quite clear that what Moyal has called a `quasi-probability distribution' is nothing more than the density matrix re-expressed in terms of the co-ordinates defined in equation (8).  These co-ordinates are essentially the mean position and momentum of what could be taken to be a cell in phase space and not the position and momentum of a particle.   In an earlier paper Hiley~\cite{bil01} proposed that it might be interesting to develop a quantum dynamics based on  an evolution of a primitive cellular structure in phase space.  This idea was further developed in~\cite{hil03} but a more rigorous approach based on the symplectic capacity has been discussed by de Gosson~\cite{mdg12}.
 We will not be discuss these important ideas further here.
 
 % Section 4.
 
 \section{The Relation to von Neumann's Primitive Idempotent.}
 
 The question I want to address here is why these apparently different starting points, namely, via (a) the characteristic function or (b) the density operator, should lead to the same time development equation (5) for $F(x,p,t)$.  To understand this connection we must go back to an  argument used by von Neumann \cite{von} to show that the Schr\"{o}dinger representation is irreducible.

von Neumann starts from the Weyl algebra which is generated by the operators $U(\alpha)=e^{i\alpha {\hat p}}$ and $V(\beta)=e^{i\beta{\hat x}}$.   These have the multiplication rules given by
\begin{eqnarray}
U(\alpha)U(\beta)=U(\alpha + \beta)  \hspace{0.5cm}  V(\alpha)V(\beta)=V(\alpha + \beta) \nonumber
\end{eqnarray}
and
\begin{eqnarray}
U(\alpha)V(\beta) = e^{i\alpha\beta}V(\beta)U(\alpha)
\end{eqnarray}									%eqn (10)
Now any quantum operator can be written symbolically in the form
\begin{eqnarray}
A=\int a(\alpha, \beta)S(\alpha, \beta)d\alpha d\beta
\end{eqnarray}									%eqn (11)
where $S(\alpha,\beta)$ is the symmetric form 
\begin{eqnarray}
S(\alpha,\beta)=e^{-i\alpha\beta/2}U(\alpha)V(\beta)=e^{i\alpha\beta/2}V(\beta)U(\alpha)  \nonumber
\end{eqnarray}
In order to find the irreducible representation of an algebra, we need to find the primitive idempotents.  To this end von Neumann chose the operator
\begin{eqnarray}
A=\int e^{-\alpha^{2}/4-\beta^{2}/4}S(\alpha ,\beta)d\alpha d\beta
\end{eqnarray}									%eqn (12)
After a little work, it is possible to show that this is, indeed, an idempotent operator as $A^{2}=2\pi A$.  What is more it can be shown that
\begin{eqnarray}
AS(\alpha, \beta)A=2\pi e^{-\alpha^{2}/4-\beta^{2}/4}A
\end{eqnarray}
which is the condition for primitivity.

The key question now is to ask how all this is related to Moyal's characteristic function.  The answer lies in some work I did for an entirely different problem. The question that concerned me was why it was possible to easily define {\em algebraically}  orthogonal spinors but not symplectic spinors \cite{hil01a} (spinors of the Heisenberg group).  The significance of this remark hinges around the fact that algebraic spinors are minimal left ideals and minimal left ideals are generated by primitive idempotents.  The orthogonal Clifford algebra, being non-nilpotent contains many idempotents that are easy to find and therefore the spinors are easy to construct \cite{fre80}. 

On the other hand the Heisenberg algebra, being nilpotent, did not contain any idempotents and therefore there seemed to be no simple way to generate the left ideals. However in the boson algebra, which can be obtained from the Heisenberg algebra using the Bargmann transformation,
we add a physically very significant primitive idempotent, namely, the projector on to the vacuum $\Omega=|0\rangle\langle 0|$.  Frescura and Hiley \cite{fre84} and Hiley \cite{hil01a}  have shown how this idempotent can be transformed back into an extension of the Heisenberg algebra and it is in this extended algebra that one can construct the symplectic spinors.

All of this  work  suggests that in the Weyl algebra we can define formally an idempotent which enables us to construct the von Neumann idempotent.  To this end let us introduce the idempotent 
\begin{eqnarray}
N(\alpha,\beta) = V(\beta)\Omega U(\alpha) 
\end{eqnarray}								%eqn (14)
That this is an idempotent follows if we introduce the annihilation and creation operators, ${\hat a}$ and ${\hat a}^{\dag}$.  Then
\begin{eqnarray}
U(\alpha)V(\beta)=e^{i\alpha\beta/2}e^{i(\alpha{\hat p}+\beta{\hat x})}=
e^{i\alpha\beta/2}e^{-|\gamma|^{2}/2}e^{i\gamma {\hat a}^{\dag}/2}e^{i\gamma^{*}{\hat a}/2 }\nonumber
\end{eqnarray}
where we have written $\gamma=(\alpha+i\beta)/2$.  After a little work, we find
\begin{eqnarray}
N^{2}=e^{i\alpha\beta/2}e^{-|\gamma|^{2}/2}N\nonumber
\end{eqnarray}
To show that this is indeed the von Neumann idempotent defined in equation (12) let us form
\begin{eqnarray}
\langle x|N(\alpha,\beta)|p\rangle&=&\int \langle x|e^{i\beta{\hat x}/2}|x^{\prime}\rangle\langle x^{\prime}|\Omega|p^{\prime}\rangle\langle p^{\prime} |e^{i\alpha{\hat p}/2}|p\rangle dx^{\prime}dp^{\prime}\nonumber\\
&=& \langle x |\Omega|p\rangle e^{i(\alpha p+\beta x)}.
\end{eqnarray}									%eqn (15)
Now
\begin{eqnarray}
\langle x|\Omega|p\rangle =\langle x|0\rangle\langle 0|p\rangle = e^{-a^{2}x^{2}/2}e^{-b^{2}p^{2}/2}\nonumber
\end{eqnarray}
where we have used the ground state wave functions, $a$ and $b$ are constants.  We have ignored normalisation factors as they do not affect our argument.   Now let us form
\begin{eqnarray}
n(\alpha,\beta)=\int \langle x|N(\alpha,\beta)|p\rangle dxdp=\int e^{-a^{2}x^{2}/2}e^{i\beta x/2}dx\int e^{-b^{2}p^{2}/2}e^{i\alpha p/2}dp.
\nonumber
\end{eqnarray}
Evaluating the integrals gives
\begin{eqnarray}
n(\alpha,\beta)=abe^{-\frac{\beta^{2}}{4a^{2}}-\frac{\alpha^{2}}{4b^{2}}}.
\end{eqnarray}
The von Neumann idempotent,  equation (11) in our notation now takes the form
\begin{eqnarray}
N(\alpha,\beta)=\int n(\alpha,\beta)S(\alpha,\beta)d\alpha d\beta.\nonumber
\end{eqnarray}										
Using the expression in equation (16) and writing $\alpha^{\prime 2}=\alpha^{2}/a^{2}$ and $\beta^{\prime 2}/b^{2}$ we have
\begin{eqnarray}
N(\alpha,\beta)=ab\int e^{-\alpha^{\prime 2}/4-\beta^{\prime 2}/4}S(\alpha,\beta)d\alpha d\beta
\end{eqnarray}									%eqn (17)
which is identical to the von Neumann idempotent except for  an unimportant constant factor.

Having identified the idempotent, all that is left to do is to show how an expression identical to what Moyal calls a characteristic function arises from this idempotent.  To do this let us introduce a complete set $|\psi_{n}\rangle$.  Then
\begin{eqnarray}
N(\alpha,\beta)=\sum _{n,n^{\prime}}e^{i\beta{\hat x}}|\psi_{n}\rangle\langle\psi_{n}|\Omega|\psi_{n^{\prime}}\rangle \langle\psi_{n^{\prime}}|e^{i\alpha{\hat p}}\nonumber
\end{eqnarray}
from which we can form
\begin{eqnarray}
\langle x^{\prime}|N(\alpha,\beta)|x\rangle=\sum_{n,n^{\prime}}C_{nn^{\prime}}
e^{i\beta x^{\prime}}\psi_{n}(x^{\prime})\psi^{*}_{n^{\prime}}(x-\alpha).
\end{eqnarray}									%eqn (18)
This indeed is a density matrix for a mixed state.  In fact this is a particular example of how to construct density matrices in general from idempotents.  This particular method is in standard use in quantum optics \cite{gla}.   In this paper we are interested in pure states so we will
choose $C_{nn^{\prime}}=\delta_{nn^{\prime}}$ and restrict the sum over the remaining index to a single value, say $n$, so that we are considering a pure state with a single component $\psi_{n}(x)$.  

We can now form
\begin{eqnarray}
\int\langle x|N(\alpha,\beta)|x\rangle=\int \psi^{*}_{n}(x-\alpha)e^{i\beta x}
\psi_{n}(x)dx.
\end{eqnarray}									%eqn (19)
If we now change variables and write $x\rightarrow y+\alpha/2$ then this equation becomes 
\begin{eqnarray}
\int\langle x|N(\alpha,\beta)|x\rangle dx =e^{i\alpha\beta/2}\int\psi^{*}_{n}(y-\alpha/2)e^{i\beta y}\psi_{n}(y+\alpha/2)dy.
\end{eqnarray}									%eqn (20)
Apart from an unimportant constant phase factor, we see that this is identical to equation (3) for the characteristic function introduced by Moyal.

\section{Conclusions.}

We have shown here that Moyal's approach which starts from the characteristic function is mathematically identical to an approach starting from the primitive idempotent that was used by von Neumann to establish the irreducibility of the Schr\"{o}dinger representation.  A different way of reaching the same conclusion can be found in Hiley~\cite{bh12}.  Thus Moyal's method  emerges from the very heart of the quantum formalism.  This confirms Baker's \cite{bak} conclusion  that the Wigner distribution is simply  the density matrix expressed in terms of a special representation based on the mean position of a pair of points in phase space.  When the approach outlined above is generalised to a six-dimensional phase space, as can easily be done, we see that the density matrix describes a cellular structure in a continuous phase space.  Thus we can think of the Wigner-Moyal approach as constructing a quantum phase space whose points are {\em subsets} of classical phase space \cite{gos1}. 

That this is a quantum phase space becomes clear from the following argument.  Consider the two operators ${\hat X}=({\hat x}^{\prime}+{\hat x})/2$ and ${\hat P}=({\hat p}^{\prime}+{\hat p})/2$.  Since $(x,p)$ and $(x^{\prime},p^{\prime})$ are conjugate points,  $[{\hat X},{\hat P}]=0$ because ${\hat x}$ and ${\hat p}$ operate on $\psi(x)$ so that $[{\hat x},{\hat p}]=i$, whereas ${\hat x}^{\prime}$ and ${\hat p}^{\prime}$ operate on $\psi^{*}(x)$ in which case $[{\hat x}^{\prime},{\hat p}^{\prime}]=-i$.   Thus the quantum phase space is constructed using the simultaneous eigenvalues of the mean operators ${\hat X}$ and ${\hat P}$.  Thus it is not correct to consider the labels $(x,p)$ in $F(x,p,t)$ as the co-ordinates of the position and momentum of a particle.

The underlying cell structure has been hinted at before,  for example, Baker \cite{bak} argued that the Wigner-Moyal formalism introduces a kind of ``smeared-out'' projection operator for a region in phase space.  However recently de Gosson \cite{gos1} \cite{gos2} has pointed out that there is a rich topological structure underlying symplectic spaces as demonstrated by Gromov's \cite{gro} ``no squeezing theorem''.  This theorem  shows that there are areas of phase space involving pairs of conjugate co-ordinates that cannot be reduced in size under classical symplectomorphisms, a kind of classical harbinger of the uncertainty principle.  Quantum mechanics introduces minimum values for this area and it is these ``quantum blobs'' that can be discussed formally in terms of the notion of a symplectic capacity~\cite{mdg12}.  Further work along these lines has been reported by Dennis, de Gosson and Hiley \cite{gdmgbh14}.

As a final remark I would like to connect the ideas outlined in this paper to the more radical ideas I have been perusing elsewhere \cite{hil03}\cite{boh06}.
This approach takes a non-commutative algebraic structure as basic and then abstracts the properties of any underlying phase space that is consistent with this algebraic structure.  For quantum mechanics, of course, this algebra is the Heisenberg algebra and its extensions.  In one approach, we follow the general ideas of Gelfand \cite{dem} and construct the points of this space from elements of the primitive idempotents. Here we use the von Neumann's idempotent to link to the Moyal approach.  If we consider this idempotent as characterising a point on the quantum phase space then we see that it has a Gaussian distribution and, of course, is a formal expression of what Baker \cite{bak} calls a ``smeared-out'' projection operator.  A more detailed account of this deeper connection with an underlying topology will be discussed elsewhere.


\begin{thebibliography}{99}

\bibitem{wig} E. Wigner, {\em Phys. Rev.}, {\bf 40}, 749-59, (1932).

\bibitem{moy} J. E. Moyal, Quantum Mechanics as a Statistical Theory,{\em Proc. Cam. Phil. Soc.}, {\bf 45}, 99-123. (1949).  

\bibitem{tak} T. Takabayasi, {\em Prog.  Theor. Phys.,} {\bf 11}, 341-74, (1954).

\bibitem{boh81}  D. Bohm and B. J. Hiley,  On a Quantum Algebraic Approach to a Generalised Phase Space,  {\em Found. Phys}., {\bf 11}, 179-203, (1981). 

\bibitem{boh83}  D. Bohm and B. J. Hiley,    Relativistic Phase Space Arising out of the Dirac Algebra,  in {\em Old and New Questions in Physics and Theoretical Biology}, Ed., A. van der Merwe,  pp. 67-76,  Plenum, (1983)

\bibitem{von} J. v. Neumann, Die Eindeutigkeit der Schr\"{o}dingerschen Operatoren, {\em Math. Ann.}, {\bf 104}, 570-87, (1931)

\bibitem{bak}  G. A. Baker, Jr., Formulation of Quantum Mechanics Based on the Quasi-Probability Distribution Induced on Phase Space, {\em Phys. Rev.}, {\bf 109}, 2198-2206, (1958).

\bibitem{bro}  M. R. Brown, and B. J. Hiley, Schr\"{o}dinger revisited: an algebraic approach.  {em quant-ph/0005026}.

 \bibitem{mor} W. Moran and J. H. Manton, in {\em Computational Noncommutative Algebra and Applications}, ed. J. S. Byrnes, pp. 339-62, NATO Science Series, Kluwer Academic, (2004).

  \bibitem{hil01} B. J. Hiley,  Towards a Dynamics of Moments: The Role of Algebraic Deformation and Inequivalent Vacuum States, {\em Correlations}, ed. K. G. Bowden, Proc. ANPA {\bf23}, 104-134, (2001).

 \bibitem{hil03} B. J. Hiley, Phase Space Descriptions of Quantum Phenomena, Proc. Int. Conf. Quantum Theory, {\em Proc. Int. Conf. Quantum Theory: Reconsideration of Foundations},  {\bf 2}, 267-86, ed. Khrennikov, A., V\"{a}xj\"{o} University Press, V\"{a}xj\"{o}, Sweden, (2003).


\bibitem{hil01a}  B. J. Hiley, A Note on the Role of Idempotents in the Extended Heisenberg Algebra, {\em Proc.  Int, Meeting, ANPA}  {\bf 22}, 107-121,Cambridge, (2001a).

\bibitem{mdg12} de Gosson, M., Quantum Blobs, {\em Found. of Phys.} {\bf 43} (2012) 1-18.

\bibitem{fre80} F. A. M. Frescura and B. J. Hiley,  The Implicate Order, Algebras, and the Spinor,  {\em Found. Phys.}, {\bf 10},  7-31, (1980) .

\bibitem{fre84} F. A. M. Frescura and B. J. Hiley, Algebras, Quantum Theory and Pre-Space,  {\em Revista Brasilera de Fisica, Volume Especial,  Os 70 anos de Mario Schonberg},  49-86, (1984).

\bibitem{gla} R. J. Glauber, Quantum Optics and Electronics in {\em  Les Houches Lectures 1964}, ed De Witt et al, p. 139, Gordon and Breach, New York, 1965.

\bibitem{bh12} Hiley, B.J., On the Relationship between the Moyal Algebra and the Quantum Operator Algebra of von Neumann, arXiv 1211.2098.

\bibitem{gos1}M. de Gosson, Uncertainty Principle, Phase Space Ellipsoids and Weyl Calculus, in {\em Operator Theory: Advances and Applications}, {\bf 164}, 121-132, Birkh\"{a}user, Basel, Switzerland 2006.

\bibitem{gos2} M. de Gosson , Phase space quantization and the uncertainty principle. {\em Phys. Lett.}, {\bf A 317}, (2003), 365-369.

\bibitem{gdmgbh14}  Dennis, G., de Gosson, M. and Hiley, B. J.,  "Fermi's ansatz and Bohm's quantum potential."  {\em Physics Letters}  {\bf A 378.32} (2014): 2363-2366.

\bibitem{gro} M. Gromov, Pseudoholomorphic curves in symplectic manifolds, {\em Invent. Math.}, {\bf 82} 307-47, 1985.

\bibitem{boh06}  D. Bohm, P. G. Davies and B. J. Hiley,  Algebraic Quantum Mechanics and Pre-geometry,  in AIP Conference Proceedings, {\bf 810}, {\em Quantum Theory: Reconsideration of Foundations--3, V\"{a}xj\"{o}, Sweden, 2005},  ed. Adenier, G.,  Khrennikov, A.,  Nieuwenhuizen, Theo., pp. 314-324 AIP, New York, 2006.

\bibitem{dem} Demaret, J., Heller, M. and Lambert,  D., Local and Global Properties of the world,   {\em Foundations of Science}, {\bf 2}, 137-176, (1997).

\end{thebibliography}
\end{document}